\documentclass[aps,prl,twocolumn,letterpaper,superscriptaddress,showpacs]{revtex4}
\usepackage{amsmath, amsthm}
\usepackage{keyval}%
\usepackage{graphicx,latexsym}
\usepackage{dcolumn}
\usepackage{bm}
\usepackage{color}
\usepackage{url}
\usepackage{CJK}


\begin{document}

\title{Itinerancy enhanced quantum fluctuation of magnetic moments in iron-based superconductors}

\begin{CJK*}{UTF8}{}

\author{Yu-Ting Tam (\CJKfamily{bsmi}譚谭宇婷)}
\affiliation{State Key Laboratory of Optoelectronic Materials and Technologies, School of Physics and Engineering, Sun Yat-Sen University, Guangzhou 510275, China}
\affiliation{CMPMSD, Brookhaven National Laboratory, Upton, NY 11973-5000,U.S.A.}

\author{Dao-Xin Yao (\CJKfamily{bsmi}姚道新)}
\affiliation{State Key Laboratory of Optoelectronic Materials and Technologies, School of Physics and Engineering, Sun Yat-Sen University, Guangzhou 510275, China}

\author{Wei Ku (\CJKfamily{bsmi}顧威)}
\altaffiliation{corresponding email: weiku@bnl.gov}
\affiliation{CMPMSD, Brookhaven National Laboratory, Upton, NY 11973-5000,U.S.A.}
\affiliation{Physics Department, State University of New York, Stony Brook, New York 11790, USA}

\date{\today}

\begin{abstract}
We investigate the influence of itinerant carriers on dynamics and fluctuation of local moments in Fe-based superconductors, via linear spin-wave analysis of a spin-fermion model containing both itinerant and local degrees of freedom.
Surprisingly against the common lore, instead of enhancing the ($\pi$,0) order, itinerant carriers with well nested Fermi surfaces is found to induce significant amount of \textit{spatial} and temporal quantum fluctuation that leads to the observed small ordered moment.
Interestingly, the underlying mechanism is shown to be intra-pocket nesting-associated long-range coupling, rather than the previously believed ferromagnetic double-exchange effect.
This challenges the validity of ferromagnetically compensated first-neighbor coupling reported from short-range fitting to the experimental dispersion, which turns out to result instead from the ferro-orbital order that is also found instrumental in stabilizing the magnetic order.
\end{abstract}

\maketitle
\end{CJK*}

In the decades-long quest to understand high-temperature superconductivity, antiferromagnetic (AFM) correlation has been one of the primary research focuses, due to its frequent proximity to superconductivity in the phase diagram.
The newly discovered Fe-based superconductors (Fe-SC) add another interesting example to such proximity: the parent compounds next to the superconducting phase possess a ($\pi$,0) C-type antiferromagnetic (C-AFM) correlation, antiferromagnetic along the $x$-direction and ferromagnetic along the $y$-direction.
This C-AFM is further accompanied by a strong coupling to a ferro-orbital (FO)~\cite{Lee09,Kruger,Singh} (or nematic~\cite{Fernanda}) and structural correlation that breaks the lattice rotational symmetry and suppresses fluctuation to the (0,$\pi$) C-AFM correlation.
Upon doping, the C-AFM correlation no longer condenses into a long-range order, but still persists at short range in the superconducting regime~\cite{Dai}, suggesting its importance in the underlying electronic structure that hosts superconductivity.

Interestingly, even in the undoped parent compounds, the C-AFM order presents an unusual characteristic: the ordered magnetic moment is much smaller than the local moment.
Indeed, Table~\ref{table1} that summarizes the experimentally estimated moments at low temperature, shows a systematic large reduction of ordered moment from the local moment across all families of Fe-SC.
That is, the local moment fluctuates strongly such that only about half of it remains ordered.
Such a systematic, strong quantum fluctuation is obviously physically significant, but poses a serious challenge to the proper understanding of the magnetism.
Currently, this fluctuation is attributed to strong temporal fluctuation from studies~\cite{G.Kotliar_BaFeAs,K.Held_LDA+DMFT,RValenti} employing dynamical mean-field approximation.
However, it is unclear whether this picture is handicapped by the intrinsic limitation of such a mean-field approximation that does not allow spatial fluctuation.

\begin{table}
\caption{\label{table1}
Large difference in local and ordered moments. \\
$\dagger$ theoretical estimation in the lack of the experimental value}
\begin{ruledtabular}
\begin{tabular}{lcr}
System &local moment$(\mu_B)$ &ordered moment$(\mu_B)$ \\
\hline
LaFeAsO & $1.1\sim2.4^\dagger$ \cite{02LaOFeAs_local}& 0.36 \cite{06_de_la_Cruz_et_al}\\
CeFeAsO & 1.3 \cite{01.local_moment_PhysRevB.85.220503} & 0.8 \cite{25.CeFeAsOnmat2315}\\
PrFeAsO & 1.3 \cite{11.local_moment_Gretarsson_et_al} & 0.53 \cite{19PrFeAsO_order}\\
FeTe    & $2\sim3$ \cite{03.FeTe3_PhysRevLett.107.216403} & $>2$ \cite{08.FeTe_T.J.Liu_Nature_Materials}\\
$\textnormal{BaFe}_2\textnormal{As}_2$ & 1.3 \cite{11.local_moment_Gretarsson_et_al} & 0.87\cite{BaFeAs_order}\\
$\textnormal{SrFe}_2\textnormal{AS}_2$ &2.1 \cite{01.local_moment_PhysRevB.85.220503} & 0.94 \cite{16.SrFeAsorderPhysRevB.78.140504}\\
\end{tabular}
\end{ruledtabular}
\end{table}

This issue is further complicated by the presence of both large local moments and the low-energy itinerant carriers in the system.
Indeed, the Fermi surface was found~\cite{Hong1,Hong2} to consist of electron and hole pockets that are approximately nested by $q=(\pi,0)$.
The presence of both itinerant carriers and local moments renders the knowledge obtained from pure local pictures (e.g. Heisenberg model) and the pure itinerant pictures (e.g. nesting of the Fermi surface) fundamentally inadequate.
Recently, it was proposed~\cite{Lv} that the itinerant carriers introduces a strong \textit{ferromagnetic} correlation along the $y$-direction (so-called double exchange effect) that stabilizes the C-AFM.
Another study~\cite{Two-fluid_Weng}, however, attributed stability of C-AFM to the assistance of the \textit{antiferromagnetic} nesting of the Fermi surface.
Obviously, it is important and timely to clarify how the itinerant carriers modulate the dynamics and fluctuation of the local moments, and to reveal the dominant magnetic mechanism and the role of the Fermi surface nesting.

In this Letter, we investigate the influence of the itinerant carriers on the dynamics and fluctuation of the local moments, using a spin-fermion model with ferromagnetic Hund's coupling.
Surprisingly, upon integrating out the itinerant degree of freedom within a linear spin-wave analysis, we found that itinerant carriers with well nested Fermi surface, instead of enhancing the C-AFM order, actually induce large \textit{spatial} and temporal quantum fluctuation that reduces the ordered moment significantly.
Interesting, the itinerancy induced renormalization of long-range magnetic couplings gives a strong \textit{antiferromagnetic} first-neighboring coupling that tends to destabilize the C-AFM order.
This is opposite to the ferromagnetic double-exchange effect previously proposed~\cite{Lv} to explain the seemingly ferromagnetically compensated first-neighbor coupling~\cite{CaFeAs_nphys1336} from short-range fitting of the experimental dispersion.
Our result challenges the validity of such fitting, and demonstrates the true origin being the FO order instead, which is found also instrumental in stabilizing the C-AFM order.
Our study not only explains the strong moment-fluctuation in Fe-SCs, but also advocates future investigations of itinerant-local interplay in strongly correlated materials.

The simplest model to describe systems consisting of local and itinerant degrees of freedom is the spin-fermion model~\cite{Yin,Lv,Weng,Dagotto}:
\begin{eqnarray}
\label{eq:eqn1}
\mathcal {H} &=& J_1\sum_{<i,i^\prime>}\vec{S}_i \cdot \vec{S}_{i^\prime} + J_2\sum_{<<i,i^\prime>>}\vec{S}_i \cdot \vec{S}_{i^\prime} \nonumber\\
 &-& J_H\sum_{i,m,s s^\prime}\vec{S}_i \cdot c_{i m s}^\dag \vec{\sigma}_{s s^\prime}c_{i m s^\prime}\\
 &-&\sum_{i i^\prime,m m^\prime, s}t_{i i^\prime}^{m m^\prime} c_{i m s}^\dag c_{i^\prime m^\prime s} + \sum_{i,m, s}(\epsilon\eta_m - \mu) c_{i m s}^\dag c_{i m s},\nonumber
\end{eqnarray}
in which the local moments $S_i$ at site $i$ couples antiferromagnetically to its first and second neighbors via $J_1$ and $J_2$, and couples ferromagnetically to the itinerant carriers via Hund's coupling $J_H$.
(Here $\vec{\sigma}_{ss^\prime}$ is the usual Pauli matrices.)
The itinerant carrier $c_{i m s}^\dag$ of orbital $m$ and spin $s$ at site $i$ hops with parameter $t_{ii^\prime}$.
Since Eq.~\ref{eq:eqn1} lacks interaction between fermions that drives the intrinsic orbital instability, we include the effect of FO order via an order parameter $\epsilon$ that shifts the $yz$/$xz$ orbital upward/downward by $\eta_m$.
The chemical potential $\mu$ is evaluated for each set of magnetic and orbital order parameters to ensures equal number of electron and hole carriers, corresponding to the undoped parent compounds.

Very often, this model is applied to incorporate only a subset of the bands of the five Fe $d$-bands as itinerant carriers, having in mind that the rest of the bands constitute the local moment~\cite{Yin,Lv,Weng,Dagotto}, consistent with the notion of orbital selective Mott transition~\cite{deMedici}.
However, such clean separation is not the only way in which the spin-fermion model can be justified or utilized.
Generally speaking, the physics to be captured in the model is how the kinetic energy of the low-energy carriers interplay with the potential energy of local moments that are low-energy in the spin (particle-hole) channel but high-energy in the one-particle spectral function. 
In line with this spirit, we include all five low-energy Fe-As hybrid $d$-orbitals in order to capture the realistic Fermi surface, knowing that only the low-energy portion of the carriers will contribute to the renormalization of magnetic coupling between local moments.
(Analogously, Hubbard's treatment of thermal fluctuation in itinerant magnetism for Fe and Ni~\cite{Hubbard2,Hubbard3} also include the full itinerant bands, with $S$ termed "exchange field".)
Specifically, we obtain first a 10-band 3D Hamiltonian from density functional theory, represented by the low-energy Wannier function~\cite{Ku2002}.
We then reduce it to a 5-band 2D Hamiltonian in the pseudo-momentum space via the local gauge transformation~\cite{supp,Lin2014,Lee_Wen} and apply it to Eq.~\ref{eq:eqn1}.

Since we are mostly interested in the generic \textit{qualitative} trends how the itinerant carriers modify the dynamics and fluctuation of the local moments in all the FeSC families, we choose the prototypical LaOFeAs for the illustration below.
We follow the procedure of Ref.~\cite{Lv} to integrate out the itinerant carriers up to the 2nd order in $J_H/S$ within the linear spin-wave theory (assisted by Holstein-Primakoff transformation, as usual)~\cite{supp} while fixing $S=1$ as a representative case.
A discrete 500x500 momentum mesh and a thermal broadening of 100meV are used to ensure a good convergence and to avoid material-specific features~\cite{broadening}.
The resulting renormalized spin-wave Hamiltonian (in unit of $S^2$)
\begin{equation}
\label{eq:eqn2}
\mathcal {H^{SW}} = \sum_q A_q(a_q^\dag a_q +a_{-q}a_{-q}^\dag) + B_q(a_q^\dag a_{-q}^\dag + a_{-q} a_q)
\end{equation}
corresponds to a Heisenberg Hamiltonian of the local moments with renormalized long-range coupling $\tilde{J}_{ii^\prime}$
\begin{equation}
\label{eq:eqn3}
\mathcal {\tilde{H}} =  \sum_{i,i^\prime} \tilde{J}_{ii^\prime} S_i \cdot S_{i^\prime},
\end{equation}
and gives the spin-wave dispersion and fluctuation of the ordered moment~\cite{supp}.
All the results below correspond to bare couplings of the local moment $J_1=19$meV and $J_2=13$meV that, after renormalization, give the correct experimental spin-wave bandwidth
($\sim 200$meV~\cite{CaFeAs_nphys1336,BaFe2As2_spin_wave_PhysRevB.86.140403,LaFeAsO_spinwave_3PhysRevB.87.140509,LiFeAs}) and a sufficiently stable C-AFM phase.

\begin{figure}[t]
\begin{center}
\resizebox*{1.0\columnwidth}{!}{\includegraphics{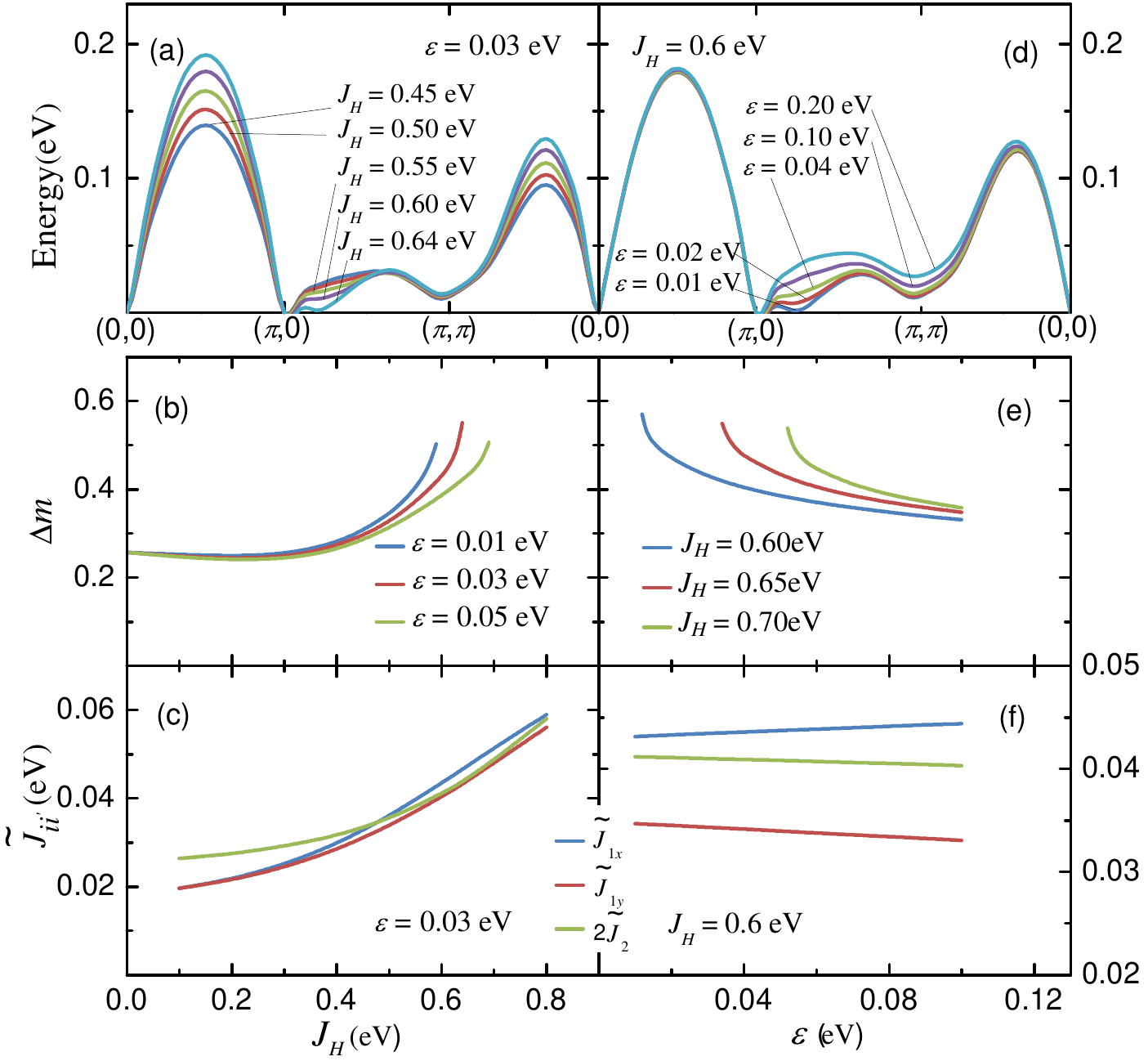}}
\end{center}
\vspace{-16pt}
\caption{(Color online) spin-wave dispersion $\omega_q$ along the high-symmetry paths (a) for a fixed FO order parameter $\epsilon$ and (d) for a fixed Hund's coupling $J_H$, (b)\&(e) reduced ordered moment $\Delta m$ for various $J_H$ and $\epsilon$, and renormalized first-neighbor couplings in $x$-direction $\tilde{J}_{1x}$, $y$-direction $\tilde{J}_{1y}$, and second neighbor coupling $\tilde{J}_2$ as functions of (c) $J_H$ and (f) $\epsilon$.}
\label{fig:fig1}
\end{figure}

To visualize the overall effects of itinerant carriers, Fig.~\ref{fig:fig1}(a) shows the resulting renormalized spin-wave dispersion for a fixed strength of FO order.
As the coupling $J_H$ to itinerant carriers increases from 0.45eV to 0.64eV, the dispersion becomes stronger along the antiferromagnetic $x$-direction [(0,0)-($\pi$,0)], while the ferromagnetic $y$-direction [($\pi$,0)-($\pi$,$\pi$)] is only weakly affected.
This indicates that the dominant physical effect is \textit{not} the ferromagnetic double exchange effect proposed previously~\cite{Lv}, but instead an enhancement of the AFM couplings.
This qualitative contrast in the physical conclusions is related to the employed fermion Fermi surface.
Focusing on cases with low density ($\sim0.1$), Ref.~\cite{Lv} used only two small electron-pockets but no hole-pockets~\cite{supp}, and thus favored nearly ferromagnetic couplings.
On the other hand, our realistic Fermi surface has compensated hole- and electron-pockets separated by ($\pi$,0) and (0,$\pi$), which promotes AFM couplings at short range.

The induced AFM coupling turns out to fluctuate significantly the C-AFM state.
Fig.~\ref{fig:fig1}(b) shows that as $J_H$ increases beyond 0.4eV, the fluctuating portion of the moment~\cite{reduced_moment}
\begin{equation}
\label{eq:eqn4}
\Delta m = 1/(8\pi^2)\int dq^2 [1-\left(B_q/A_q\right)^2]^{-1/2}-1/2
\end{equation}
grows quickly and easily exceed 0.5, half of the local moment, accounting for the experimental observation.
Since this quantum fluctuation originates from spontaneous creation of propagating magnons in a perfectly ordered state~\cite{PhysRev.87.568,Auerbach}, it is naturlaly \textit{spatial} and temporal.

More detailed microscopic understanding of this AFM coupling-induced fluctuation can be obtained from Fig.~\ref{fig:fig1}(c) that shows the renormalized first-neighboring couplings in $x$- and $y$-directions ($\tilde{J}_{1x}$ and $\tilde{J}_{1y}$), and the renormalized second-neighboring coupling ($\tilde{J}_2$), for the case of $\epsilon=30$meV as an example.
As $J_H$ grows near 0.6eV, $\tilde{J}_{1y}$ increases faster than 2$\tilde{J}_2$ and eventually approaches the latter, a situation in which the C-AFM state becomes unstable against ($\pi$,$q$)-fluctuation ($-\pi\leq q \leq \pi$)  according to the simple classical energy estimation of the short-range $J_1$-$J_2$ model.

In other words, even though itinerant carriers themselves, having an approximate ($\pi$,0) nesting, prefer C-AFM correlation, their influence on the couplings of local moments, however, gives a larger $\tilde{J}_{1y}$ that disfavors the C-AFM correlation and produces strong fluctuation.
Evidently, the previously reported itenerant nesting-enhanced susceptibility at ($\pi$,0)~\cite{Two-fluid_Weng} is insufficient to conclude a more stabile C-AFM state.
Instead, the even stronger enhancement of ($\pi$,$q$) fluctuation reveals the dominant role of itinerant carriers in reducing the C-AFM ordered moment.
Such a surprising effect merely reflects the rich interplay between the itinerant carriers and local moments in our model, and set an alarming example against the usual itinerant-only considerations (e.g. the typical normal-state nesting arguments).

\begin{table}
\caption{\label{table2}Importance of induced long-range couplings}
\begin{ruledtabular}
\begin{tabular}{lcc}
$J_H\, =\, 0.6\,$eV, & $ q =(\pi,\frac{\pi}{5})$ & $q =(\pi,\pi)$\\
$\epsilon\,=\,0.03\,$eV & full/up-to-$\tilde{J}_2$ & full/up-to-$\tilde{J}_2$\\
\hline
$A_q$ & 0.1534/0.1558 & 0.0134/0.0316   \\
$B_q$ &-0.1527/-0.1532 & 0.0022/-0.0049 \\
$w_q$ & 0.0103/0.0281 & 0.0132/0.0312  \\
\end{tabular}
\end{ruledtabular}
\end{table}

It is important to note the significant contributions of the RKKY-like long-range (power-law decaying) couplings between local moments induced by itinerant carriers~\cite{long-range}.
Indeed, besides the leading $\tilde{J}_{1x}$, $\tilde{J}_{1y}$, and $\tilde{J}_2$, the resulting renormalized coupling $\tilde{J}$ contains small but long-range components that add up to important contributions.
Table~\ref{table2} gives an example of the calculated $A_q$, $B_q$ and the spin-wave energy $w_q$, with and without the long-range couplings.
It shows that inclusion of long-range couplings can change these values by a factor of two, confirming their importance.
This is another alarming message to the common analysis of inelastic neutron scattering measurements that use only short-range couplings to fit the experimental data.
For systems with itinerant carriers and consequently long-range magnetic couplings, such fitting can be dangerously misleading about the important physical effects.

The long range of the couplings naturally enables local moments to fluctuate with extended \textit{spatial} structures.
For example, the spin-wave dispersion in Fig.~\ref{fig:fig1}(a) shows a softening in the vicinity of ($\pi$,$\pi$/5), indicating strong fluctuation approximately in period of 10 unit cells.
Our results therefore conclude a strong \textit{spatial} fluctuation in addition to the currently emphasized temporal fluctuation~\cite{G.Kotliar_BaFeAs,K.Held_LDA+DMFT,RValenti} of the magnetic moment.

\begin{figure}[t]
\begin{center}
\resizebox*{1.0\columnwidth}{!}{\includegraphics{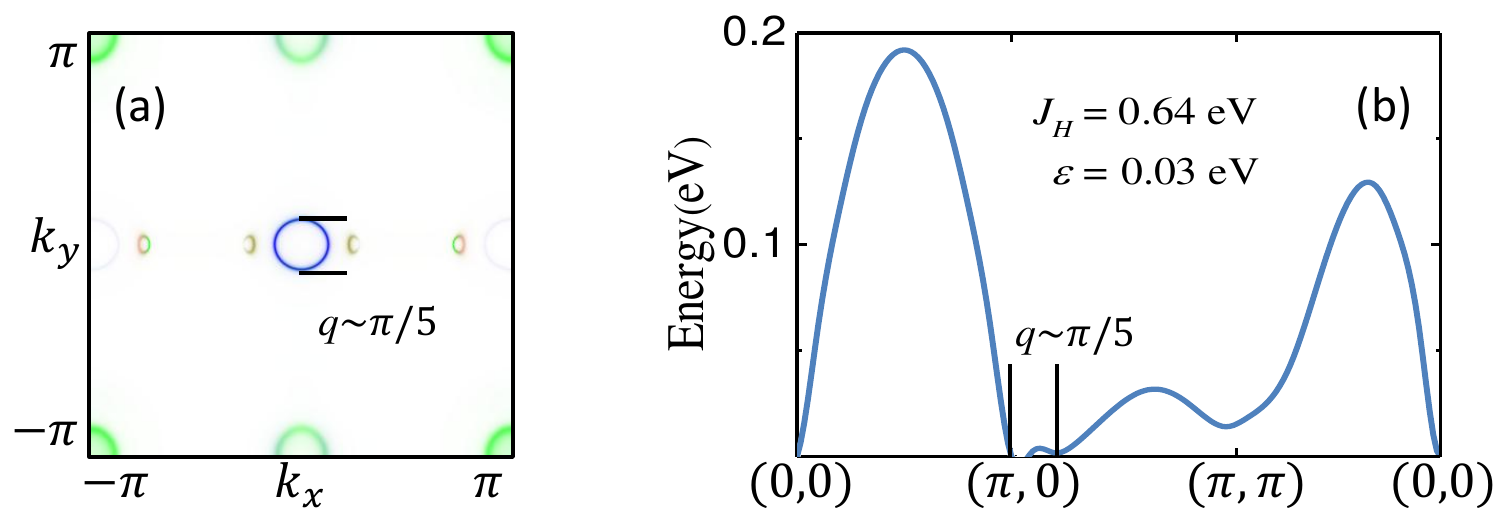}}
\end{center}
\vspace{-16pt}
\caption{(Color online) (a)Nesting of Fermi surface of a C-AFM and FO ordered state, unfolded in one-Fe Brillouin zone, and (b)its effect on the momentum region with spin-wave softening and enhanced fluctuation.}
\label{fig:fig2}
\end{figure}

The origin of the enhanced fluctuation in the vicinity of ($\pi$,$\pi$/5) can actually be associated with the Fermi surface of the \textit{ordered} state.
As an illustration, Fig~\ref{fig:fig2}(a) shows the Fermi surface modified by the C-AFM and FO order parameters, unfolded in one-Fe Brillouin zone.
One can observe an approximate intra-pocket nesting of (0, $2k_F$)$\sim$(0, $\pi$/5), equivalent to ($\pi$, $\pi$/5), matching the momentum region of the spin-wave softening and the enhanced fluctuation in Fig~\ref{fig:fig2}(b).
In essence, the short-range frustration of couplings enhances the overall spin fluctuation along ($\pi$,0) to ($\pi$,$\pi$), while the nesting of the Fermi surface of the \textit{ordered} state pinpoints a more specific momentum region with most fluctuation.

\begin{figure}[t]
\begin{center}
\resizebox*{1.0\columnwidth}{!}{\includegraphics{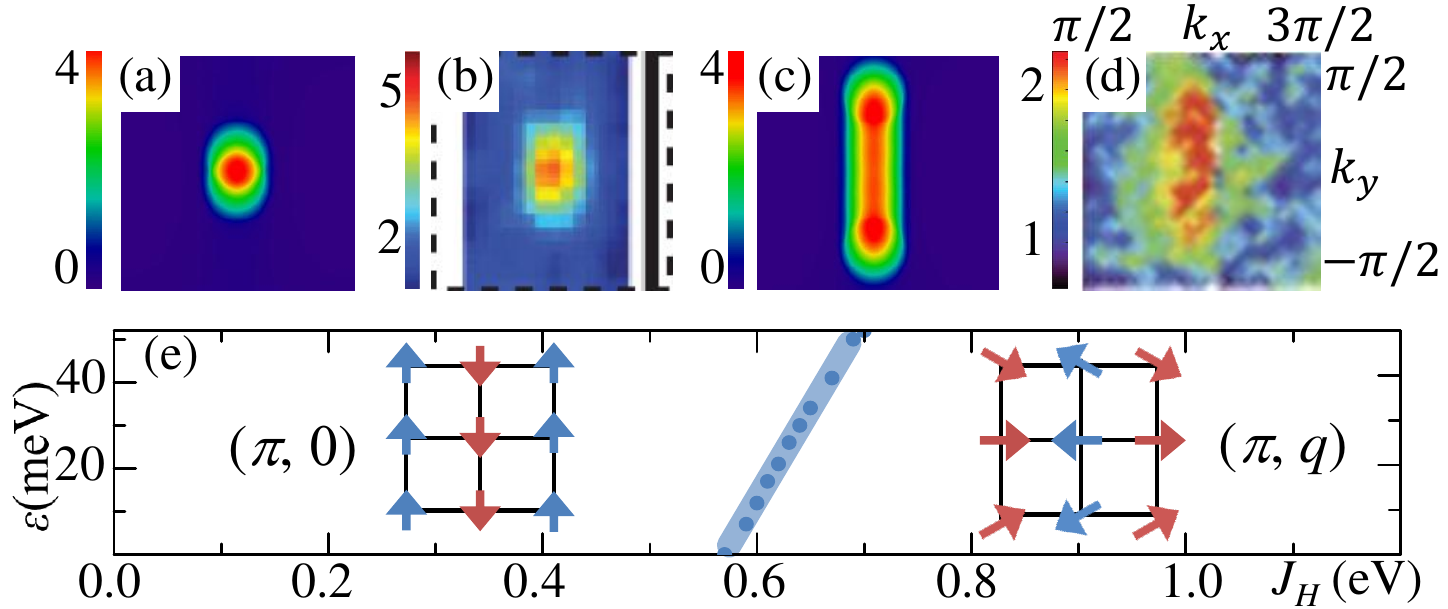}}
\end{center}
\vspace{-16pt}
\caption{(Color online)
Theoretical momentum distribution of magnetic excitation at 10$\pm 1$meV in (a) the C-AFM state ($J_H$=0.45eV,$\epsilon$=30meV) and (c) a highly fluctuating state ($J_H$=0.64eV,$\epsilon$=30meV), in 0.07$\pi$ momentum-resolution.
The latter shows enlongated ($\pi$,$q$) fluctuation, as measured in (d) spin-disordered FeTe$_{0.51}$Se$_{0.49}$~\cite{IC_excitation} when compared with (b) spin-ordered BaFe$_2$As$_2$~\cite{PhysRevB.84.054544}. (e) Phase diagram showing the stable region of the C-AFM state.
}
\label{fig:fig3}
\end{figure}

Note that the ``incommensurate'' ($\pi$,$q$) excitation has in fact been observed in several materials~\cite{Dai,IC_excitation,FeSe2,FeSe3}.
Figure~\ref{fig:fig3}(a)-(d) compare theoretical and experimental low-energy excitations of a C-AFM state and a strongly fluctuating state.
The latter clearly has a more pronounced ($\pi$,$q$) momentum distribution, confirming their direct connection to the moment fluctuation.

Having established the main physical effects of the itinerant carriers, we now illustrate the essential role of the FO order in the C-AFM state using the right panels of Fig~\ref{fig:fig1}.
Figure~\ref{fig:fig1}(d) shows clearly that a stronger FO order increases the dispersion along the path from ($\pi$,0) to ($\pi$,$\pi$), but barely changes the dispersion along the antiferromagnetic $x$-direction.
In some of the Fe-SCs, such an enhanced spin-wave dispersion along the ferromagnetic $y$-direction has in fact been observed experimentally~\cite{CaFeAs_nphys1336}, and was previously attributed to a nearly compensated $\tilde{J}_{1y}$.
This appears to suggest a large ferromagnetic double exchange effect.
However, as discussed above, our realistic Fermi surface produces largely enhanced AFM $\tilde{J}_{1x}$, $\tilde{J}_{1y}$ and $\tilde{J}_2$, and thus does not support the double exchange picture.
Indeed, Fig.~\ref{fig:fig1}(f) shows that this remains true even after introduction of the FO order:
While $\tilde{J}_{1x}$ and $\tilde{J}_{1y}$ are slightly split by the anisotropy introduced by FO, they remain strongly antiferromagnetic.

Our results in Fig~\ref{fig:fig1} thus uncovers the FO order being the true key physical effect.
More microscopically, it is the FO order induced anisotropy in the renormalized coupling $\tilde{J}$ that raises the spin-wave energy along the path from ($\pi$,0) to ($\pi$,$\pi$) and slightly reduce the fluctuation.
While the C-AFM order can also introduce the anisotropy, Fig.~\ref{fig:fig1}(a)(b) clearly indicates their different physical effects: C-AFM order enhances the dispersion along the $x$-direction, but FO order the $y$-direction.
It is now also understandable why the experimental fitting may lead to nearly compensated $\tilde{J}_{1y}$: the fitting assigns all the anisotropy to one short-range coupling, even though in reality such anisotropy is spread out to a large number of long-range couplings instead, dictated by the itinerant nature.


In fact, the FO order is found essential to the C-AFM order.
For example, in Fig.~\ref{fig:fig1}(f) $\tilde{J}_{1y} > 2\tilde{J}_2$ suggests that the C-AFM state should already be unstable if it weren't for the FO-induced anisotropy in the couplings.
Indeed, one sees in Fig.~\ref{fig:fig1}(e) that a stronger FO order helps to reduce the moment fluctuation.
Within the realistic parameter range, our result gives a phase diagram in Fig.~\ref{fig:fig3}, in which the C-AFM order requires the FO order for $J_H > 0.57$eV.
This accounts naturally for the experimental fact across all Fe-SC families, that the FO/structural order always precedes the magnetic order~\cite{25.CeFeAsOnmat2315,phase_Rotundu,phase_Ni,phase_Kasahara}.
Of course, if the fluctuation in the extended ($\pi$,$q$) region overwhelms the aid from the FO order, the C-AFM order will still be destroyed.
This gives the simplest explanation of the absence of magnetic order in FeSe, which still presents a strong FO/structural order~\cite{phase_FeSe}.
Sure enough, spin-disordered Se-rich FeTe$_{1-x}$Se$_x$~\cite{IC_excitation,FeSe2,FeSe3} samples all present extended ($\pi$,$q$) fluctuation similar to Fig.~\ref{fig:fig3}(d).

In conclusion, we investigate the generic effects of itinerant carriers on the dynamics and fluctuation of local moments in the parent compounds of Fe-based superconductors, by integrating out the itinerant carriers that couple to the local moment via Hund's coupling.
Unexpectedly, while the ($\pi$,0)-nested itinerant carriers alone prefers a C-AFM order, they however fluctuate efficiently C-AFM ordered local moments and produce the experimentally observed large reduction of the latter.
Interestingly, the itinerant carriers induce long-range couplings between local moments with AFM first-neighboring couplings, producing strong \textit{spatial} and temporal quantum fluctuation.
This surprising finding is opposite to the common intuition of the nesting effect, and is distinct from the ferromagnetic double-exchange physics proposed previously to explain the seemingly ferromagnetic compensated first-neighbor coupling from short-range fitting of measured spin-wave dispersion.
Our result challenges the validity of such fitting and demonstrates that such dispersion results naturally from the FO order, which is also essential in stabilizing the C-AFM.
These new insights illustrate the generic rich interplay between itinerant and localized degrees of freedom in many-body systems, and advocate further systematic investigation of transport, superconductivity, and other fluctuation dominant phenomena in these systems.

We thank Fan Yang and Hai-Qing Lin for helpful discussion, and Beijing Computational Science Center for its warm hospitality in hosting part of the research activity.
WK acknowledges support from U.S. Department of Energy, Office of Basic Energy Science, Contract No. DE-AC02-98CH10886.
YT and DXY acknowledge support from NBRPC-2012CB821400, NSFC-11275279, SRFDP-20110171110026,and NCET-11-0547.

\bibliography{moment_flutuation_r2v2}

\end{document}